\documentclass[A4paper, 12pt]{article}
\usepackage{grffile}
\usepackage{lineno}
\usepackage{lipsum}
\newcommand\numberthis{\addtocounter{equation}{1}\tag{\theequation}}
\usepackage[toc,page]{appendix}
\usepackage{epstopdf}
\usepackage[UKenglish]{isodate}
\usepackage{pgfplots}
\newtheorem{example}{Example}
\usepackage{mathtools}
\usepackage{natbib}
\usepackage{tikz}
\usepackage{graphicx}
\usepackage[toc]{appendix}
\usepackage[font={small,it}]{caption}
\usepackage{blindtext}
\usepackage{subcaption}
\usepackage[T1]{fontenc} 
\usepackage{mathtools}
\usetikzlibrary{shapes,snakes,shadings, fadings}
\usetikzlibrary{arrows}
\tikzstyle{block}=[draw opacity=0.7,line width=1.4cm]
\graphicspath{ {images/} }
\usepackage{setspace}
\usepackage{mfirstuc}
\usepackage{float}
\usepackage[utf8]{inputenc}
\usepackage{pgfplots}
\usetikzlibrary{datavisualization.formats.functions}
\usepackage{amsmath,amssymb,mathtools,bm,etoolbox}
\title{Adversarial Risk Analysis for First-Price Sealed-Bid  Auctions}
\author{Muhammad Ejaz, Chaitanya Joshi and Stephen Joe}
\date{Mar 18, 2020}

\begin{document}

\maketitle

\begin{abstract}
	Adversarial Risk Analysis (ARA) is an upcoming methodology that is considered to have advantages over the traditional decision theoretic and game theoretic approaches. ARA solutions for first-price sealed-bid (FPSB) auctions have been found but only under strong assumptions which make the model somewhat unrealistic. In this paper, we use ARA methodology to model FPSB auctions using more realistic assumptions. We define a new utility function that considers bidders' wealth, we assume a reserve price and find solutions not only for risk-neutral but also for risk-averse as well as risk-seeking bidders. We model the problem using ARA for non-strategic play and level-\textit{k} thinking solution concepts.	
\end{abstract}



\section{Introduction} \label{sec:intro}
\subsection{Decision vs Game Theoretic Approaches}
The use of auctions is from time immemorial. The first modern academic paper for \emph{first-price sealed-bid} (FPSB) auctions\footnote{Bidders place their bids in sealed envelopes and simultaneously hand over them to the auctioneer. Those envelopes are then opened and the bidder with the highest bid wins and pays the amount equal to the bid.} was presented by \cite{friedman1956competitive} and used \emph{decision theoretic} approach. Further work using decision theoretic models followed. \cite{capen1971competitive} and \cite{keefer1991development} reported that in practice, bidders do indeed use decision theory. \cite{rothkopf1994modeling} critically analysed the models available for auctions design and bidding strategy in real situations and concluded that bidders were using decision theoretic models to decide upon their bids. \cite{rothkopf2007decision} gave several arguments that for FPSB auctions, decision theory dominates game theory. \cite{engelbrecht2007regret} also use decision theoretic model for these auctions assuming risk-neutral bidders while taking into account bidder's winning and losing regret.  \cite{wang2017behavioral} developed a decision theoretic model to find the optimal bid for a bidder in these auctions and stated that risk neutral \emph{Bayes Nash equilibrium} (BNE) cannot account for a bidder's actual behaviour in bidding.

On the other hand, a large number of \emph{Bayesian game theoretic} approaches have been proposed for the FPSB auctions as well. \cite{vickrey1961counterspeculation} analysed $ n $ bidders in FPSB auctions using Bayesian game theory  with the values drawn from a uniform distribution with common support. \cite{criesmer1967toward} analysed  the BNE of FPSB auctions in which bidders’ valuations were drawn from uniform distributions  with  different  supports, while \cite{wilson1969communications} developed the first closed form equilibrium analysis with the  common value model (value of the auctioned item is same for all bidders). \cite{riley1981optimal} extended Vickrey's analysis to $ n $ bidders whose values were drawn independently and identically from a general commonly known distribution that was strictly increasing and differentiable. Then, \cite{myerson1981optimal} and \cite{milgrom1982theory}, among others, also used Bayesian game theoretic model for auctions. \cite{cox1982theory} and \cite{cox1982auction} generalised the Vickrey's model to the case of risk-averse bidders who have the \emph{constant relative risk aversion} (CRRA) utility functions. The CRRA utility function is defined later in Subsection \ref{UF}. \cite{maskin2000asymmetric} analysed FPSB auctions assuming the valuations of each bidder is drawn from commonly known different distributions. \cite{goeree2002quantal, bajari2005structural, campo2011semiparametric, gentry2015existence, li2017hidden} among others also used Bayesian game theoretic models for auctions.

However, both these approaches have their drawbacks. While a decision theoretic approach does not require the \emph{common knowledge} assumption, it does assume that the other bidders are non-strategic. Assuming  non-strategic bidders may be unrealistic because bidders may often be strategic. On the other hand, using a Bayesian game theoretic model requires a strong common knowledge assumption that all bidders who are considered to be strategic, draw their valuation for the auctioned item from a commonly known distribution. The common knowledge assumption can also be unrealistic because the bidders may draw their valuation from different distribution(s) and the distribution used by one bidder is usually not commonly known to others. In fact, often the bidders will try and keep their information secret so as to gain competitive advantage. Also, finding a BNE becomes increasingly difficult as the games get more realistic (more complex) and often it may be that a unique BNE solution to a given game does not exist.

\subsection{Adversarial Risk Analysis} \label{ara}
To overcome the shortcomings of both the decision theoretic as well as the Bayesian game theoretic approaches, \cite{rios2009adversarial} introduced an approach called the \emph{adversarial risk analysis} (ARA) to model the decision making problems in the presence of an intelligent adversary such as those encountered in cyber-security, counter-terrorism, war, politics, auctions, etc. ARA is a Bayesian approach because subjective distributions are used to model the uncertainties about the outcomes and about the unknown preferences, beliefs and capabilities of intelligent adversaries. However, unlike the Bayesian game theory, ARA does not require that these subjective distributions be the same for all the players and that they be commonly known. Another important difference to Bayesian game theory is that ARA aims to solve the decision making problem for just one of the player's and does not aim to find an equilibrium solution for all the players. For this reason, finding an ARA solution is relatively less difficult even for complex problems.\\
Since its introduction, it has been used to model a variety of problems such as network routing for insurgency (\cite{wang11}), international  piracy (\cite{sevillano12}), counter-terrorism (\cite{rios12}),  autonomous social agents (\cite{esteban14}, urban security resource allocation (\cite{gil16}),  adversarial classification (\cite{naviero19}), counter-terrorist online surveillance (\cite{Gil19}), cyber-security (\cite{rios19}) and insider threat (\cite{joshi20}).\\

We will illustrate how ARA works by considering a two player game  between Brenda $ (B) $, and Charles $(C)$. Let $b$ and $c$ be the choices they make from their respective sets of actions $\mathcal{B}$ and $\mathcal{C}$. The resulting outcome $ S $ is a chance variable which takes a value $s$ from a set of possible outcomes $\mathcal{S}.$ Let $ u_B(b,c, s) $ and $ u_C(b,c,s) $ be the utility functions that determine the utilities received by Brenda and Charles respectively, given a pair of actions $ (b,c) $ and outcome $s$. Suppose that we are solving the problem for Brenda. Then, a typical objective (although, other objectives are possible too) will be to find the optimal action $b^{*}$ that maximizes her expected utility. Figure \ref{id3} represents this game in the form of a bi-agent influence diagram (BAID) where rectangles, circles and hexagons represent decision, chance  and utility nodes respectively. Figure \ref{id1} represents this game from both Brenda and Charles's point of view while Figure \ref{id4} represents this game from Brenda's point of view only where, Charles's decision node is now a chance node for Brenda because she is uncertain about his actions. 

\tikzstyle{decision} = [rectangle, draw, 
text width=1.5em, text centered, minimum height=1.9em]
\tikzstyle{line} = [draw, -latex']
\tikzstyle{chance} = [draw, circle, node distance=1cm,
minimum height=2.3em]
\tikzstyle{utility} = [draw, regular polygon, regular polygon sides=6, node distance=1cm, minimum height=1em, text width=0.9em] 

\begin{figure} [H]
	\centering
	\begin{subfigure} [t]{0.47\textwidth}
		\centering
	\begin{tikzpicture}[node distance = 2cm, auto]
	\node [chance, left color=gray!30] (e) {$ S $};
	\node [decision, above left of=e, node distance=2.5cm, fill=gray!20] (b) {$ B $};
	\node [decision, above right of=e, node distance=2.5cm] (c) {$ C $};
	\node [utility, below right of=e, node distance=2.5cm] (uc) {$ u_C $};
	\node [utility, below left of=e, node distance=2.5cm, fill=gray!20] (ub) {$ u_B $};
	\path [line] (b) -- (e);
	\path [line] (c) -- (e);
	\path [line] (b) -- (ub);
	\path [line] (c) --(uc);
	\path [line] (e) -- (ub);
	\path [line] (e) -- (uc);
	\end{tikzpicture}
	\caption{} 
	\label{id1}
\end{subfigure}
\hfill
\begin{subfigure} [t]{0.47\textwidth}
	\centering
\begin{tikzpicture}[node distance = 2cm, auto]
\node [chance, left color=gray!30] (e) {$ S $};
\node [decision, above left of=e, node distance=2.5cm, fill=gray!20] (b) {$ B $};
\node [chance, above right of=e, node distance=2.5cm] (c) {$ C $};
\node [utility, below right of=e, node distance=2.5cm] (uc) {$ u_C $};
\node [utility, below left of=e, node distance=2.5cm, fill=gray!20] (ub) {$ u_B $};
\path [line] (b) -- (e);
\path [line] (c) -- (e);
\path [line] (b) -- (ub);
\path [line] (c) --(uc);
\path [line] (e) -- (ub);
\path [line] (e) -- (uc);
\end{tikzpicture}
\caption{} 
\label{id4}
\end{subfigure}
\caption{(a) The BAID for the two player simultaneous game (b) The BAID for the two players from Brenda's perspective} \label{id3}
\end{figure}
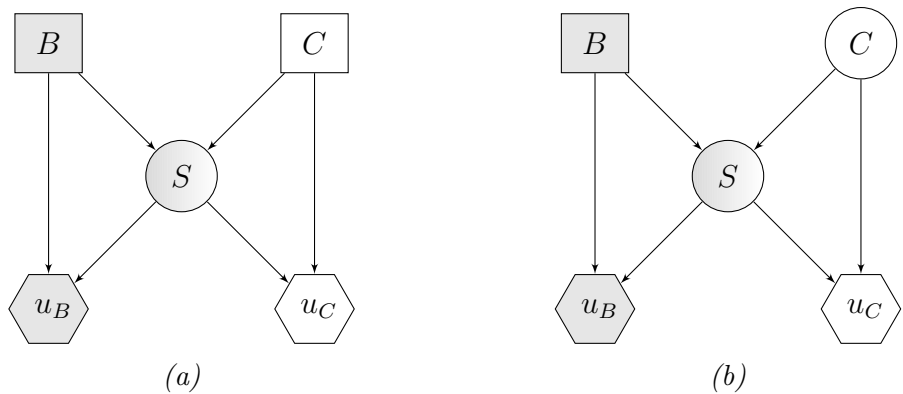

The basic idea behind finding an ARA solution is to \emph{integrate out} the uncertainties at the chance nodes and \emph{maximize} the expected utility at the decision nodes. When solving the problem for Brenda, we have two chance nodes, namely $C$ and $S$ and the sole decision node $B.$ ARA can be solved using backward induction. We can solve the problem for Brenda as follows.\\
We first find her expected utility by taking into account her uncertainty about the outcomes $S$
\begin{equation} \nonumber
\varPsi_B(b,c)=\int u_B(b,c, s)p_B(s|b,c)\, ds,
\end{equation}
where $ p_B(s|b,c) $ denotes Brenda's uncertainty in $ S $ given actions $b$ and $c$. Next, we can find her expected utility by taking into account her uncertainty about Charles's actions as
\begin{equation} \label{eu2}
\varPsi_B(b)=\int \varPsi_B(b,c)p_B(c)\, dc,
\end{equation}
where $ p_B(c) $ is Brenda's uncertainty in $C$. Finally, we can find the optimal action $ b^\ast $ that maximizes her expected utility as
\begin{equation*} \label{b}
b^\ast=\text{argmax}_{b \in \mathcal{B}}\,\varPsi_B(b).
\end{equation*}
The main challenge in the above modelling is to determine $ p_B(c) $. Brenda may elicit $ p_B(c) $ either by using her subjective beliefs or by using data on the past auction bids by Charles on similar items or by using an expert opinion. But, she could also choose to elicit $ p_B(c) $ by modelling Charles's strategic thinking process. For example, she may believe that Charles is an expected utility maximizer, just like her, and would choose the action $ c^\ast $ that maximizes his expected utility. She can aim to find $ c^\ast ,$ if Charles's utility function $ u_C(b,c,s) $ and his probabilities $ p_C(b) $ and $ p_C(s|b,c) $ are available to her. However, these quantities are typically not available to her. She can model her uncertainty about Charles's utility and his probabilities through eliciting random utility $ U_C(b,c,s) $ and random probabilities $ P_C(b) $ and $ P_C(s|b,c) $. Then, following a backward induction approach similar to her own, she can find Charles's random expected utility as
\begin{equation} \nonumber
\Psi_C(c)=\int \Psi_C(b,c)P_C(b)db,
\end{equation}
where
\begin{equation} \nonumber
\Psi_C(b,c)=\int U_C(b,c, s)P_C(s|b,c)ds.
\end{equation}
Then, she could find his random optimal action $ C^\ast $ as
\begin{equation*} \label{eu3}
C^\ast=\text{argmax}_{c \in \mathcal{C}}\Psi_C(c).
\end{equation*}
Finally, she could find her required predictive distribution $ p_B(c) $ about Charles's action $ c $ as
\begin{equation} \label{eu4}
\int_{-\infty}^{c} p_B(\varphi)d\varphi=Pr(C^\ast \leq c).
\end{equation}

We have assumed in our notation above that the sets of actions $\mathcal{B}$ and $\mathcal{C}$ and the set of outcomes $\mathcal{S}$ are open intervals or unions of open intervals that contain infinitely many points. If any of these sets were of finite cardinality then the corresponding integrals would be replaced by sums.\\
In the above analysis, we have assumed that Charles is an expected utility maximizer when assessing $ p_B(c).$ However, this probability can be assessed by assuming that Charles uses a number of alternative solution concepts. For example, Brenda can solve this problem by assuming that Charles is a non-strategic player or that he is a level-\textit{k} thinker or that he uses BNE to determine his optimal action and so on. An ARA solution can be found for each of these solution concepts. See \cite{banks2015adversarial} for details and also for how ARA can be used to solve different types of games, such as sequential games, multi-player games, etc.\\

\cite{banks2015adversarial} modelled  FPSB auctions assuming that each bidder is risk-neutral.  They derived ARA solutions assuming that the opponent has different solution concepts such as non-strategic play, minimax perspective, level-\textit{k} thinking, mirror equilibrium and BNE. However, they did not consider a bidder's wealth while defining their utility function. Also, they did not model risk-averse or risk-seeking behaviour of the bidders. They also did not consider reserve price for the auctioned item which is a common practice in FPSB auctions. In reality, however, it is well known that bidder's wealth, their risk appetite and whether the item had a reserve price or not, all play a role in determining the bid the item may attract. In this paper, we aim to model FPSB auctions considering these factors and derive ARA solutions for the non-strategic play and level-\textit{k} thinking solution concepts.

\subsection{Utility Function} \label{UF}
First we define the three risk behaviours modelled in this paper. It is important to note that in the context of auctions, \emph{risk} relates to the risk of not winning the item.  A risk-neutral bidder is defined as the one who is indifferent to risk when making a bidding decision. A risk-neutral behaviour could either be due to a rational decision making process and a person taking a calculated balanced approach or it could be an emotional preference whereby the person simply does not focus on the risk involved. Risk-averse bidders are the bidders who do not want to lose the item. Thus, they bid more aggressively than the risk-neutral bidders. On the other hand, risk-seeking bidders are the bidders who are keen to get the item at a low price. Thus, they bid less aggressively than the risk-neutral bidders. Previous literature reveals that risk aversion is an important determinant of bidders' bidding behaviour in auctions \cite[see e.g.][among others]{milgrom1982theory, maskin1984optimal,  gentry2015existence}. Specifically in FPSB auctions, a bidder does not really know about other bidders' behaviours. For this reason, risk aversion is more significant to be taken into account for these auctions \cite[see  e.g.][]{cox1988theory, kagel1995auctions, dorsey2003explaining}. \\
A commonly used utility function in the literature of auctions for risk-averse bidders is the CRRA utility function and it is so because of its computational ease (\cite{holt2002risk}). A basic CRRA utility function for a bidder having wealth $ w $ is defined as
\begin{equation} \label{uf}
u(w)=w^r, \qquad w >0,
\end{equation}
where, $ (1-r)=-wu''(w)/u'(w) $ is the coefficient of CRRA or Arrow-Pratt measure of relative risk-aversion \cite[]{pratt1964risk, arrow1965aspects}. The coefficient of CRRA measures the proportion of wealth an individual will choose to hold on a risky asset, for a given level of wealth $ w $. 
The utility function \eqref{uf} is strictly convex for $ 1<r \leq 2 $ which represents the risk-seeking behaviour, it is linear for $ r=1 $ which represents the risk-neutral behaviour and it is strictly concave for $ 0 <r <1 $ which represents the risk-averse behaviour \cite[see e.g.][for more details]{holt2002risk}.

\cite{cox1982theory, cox1985experimental,  marechal2011first}, among others, used \eqref{uf} without considering bidders' wealth and defined their utility function as
\[
u(b,v)=(v-b)^r,
\]
where, $ v $ is a bidder's true value and $ b $ is the successful bid. 

On the other hand, for example, \cite{lu2008estimating, li2017hidden}, among others, used \eqref{uf} while also considering bidders' wealth and defined their utility function in case of their successful bid as
\begin{equation} \label{ufw} 
u(b,v,w)=(w+v-b)^r,
\end{equation}
where, they assumed that all bidders have the same wealth $ w\geq 0 $.\\
However, in Section \ref{ufp} we show that this utility function is unrealistic for multiple reasons. We propose a new utility function that is more realistic and does not have the drawbacks that \eqref{ufw} has.\\

\noindent{\textbf{Normal goods:}} In Economic theory a good that has positive income elasticity is defined as a \emph{normal good}. That is, demand for a normal good rises when income increases and falls when the income falls (see, for example, \cite{goree16, hoboken13, fisher1990normal, Jeffrey2015mic}). It is an item for which a person's demand increases with increase in her wealth \cite[]{baisa2017auction}. The items that are typically considered  to be normal include consumables, but also items such as collectibles, houses, cars, jewelry, etc. The FPSB auction has been used in Japan to sell dried fish which could be considered as a normal item.\\
Thus, when bidding on a normal item, a bidder could be more risk-averse (or less risk-seeking) with increase in their wealth. That is, relative wealth and risk behaviour are linked and it is therefore not enough to model the risk behaviour by using a CRRA parameter that is not directly linked to the relative wealth. However, except for  \cite[]{baisa2017auction}, the type of goods auctioned has not specifically been taken into account when defining utility functions and CRRA parameters. 

\subsection{Contributions in this Paper} \label{Contr}
The main contributions contained in this paper are as follows:
\begin{itemize}
	\item [$\bullet$] We extend \cite{banks2015adversarial} by developing ARA solutions for non-strategic play and level-\textit{k} thinking solution concepts for a realistic case, where in,
	\begin{itemize}
		\item [$\bullet$] we consider a reserve price for the auctioned item (typically, known to each bidder in advance).
		\item [$\bullet$] we take into account bidders' wealth.
		\item [$\bullet$] we find solutions not only for risk-neutral bidders but also for risk-averse and risk-seeking bidders.
	\end{itemize}
	\item [$\bullet$] we assume that the auctioned item is normal and define a new bidder's CRRA utility function. 
	\item [$\bullet$] unlike the utility function \eqref{ufw}, where it is assumed that all bidders have same wealth, we assume that the bidders may have different wealths. 
	\item [$\bullet$] we use the CRRA parameter $ r $ and also define a new CRRA parameter $ a $ to incorporate the effect of increase in wealth on bidder's risk behaviour.  
\end{itemize}

\subsection{Structure of the Paper}
In this paper, we assume that we are finding ARA solutions for Brenda $ (B)  $ against her opponent Charles $ (C) $ in in an FPSB auction. The remainder of the paper is organised as follows. In Section \ref{ufp}, we propose our new utility function and the new CRRA parameters. In Section \ref{NSP}, we derive ARA solution for the FPSB game where Brenda assumes that Charles a is non-strategic player. In Section \ref{k}, we derive the solution where Brenda assumes that Charles is a level-\textit{k} thinker.  In both the Sections \ref{NSP} and \ref{k}, we also illustrate the ARA solutions with detailed numerical examples. Finally, in Section \ref{d}, we discuss the results we obtained in this paper and sketch ideas for further work.

\section{New Utility Function and New CRRA Parameters} \label{ufp}

\subsection{Drawbacks of utility function \textbf{\eqref{ufw}}} \label{uf0}

For the utility function \eqref{ufw}, the wealth $ w $ has been defined in many ways in the auction literature. For example, $ w \geq 0 $ \cite[]{lu2008estimating, li2017hidden}, $ 0 \leq w \leq \bar w $, where $ \bar w $ is the upper support of wealth among $ n $ bidders \cite[]{gentry2015existence} and $ w>\varrho $, where $ \varrho $ is the entry cost of auction \cite[]{li2015auctions}. However, none of these constraints incorporate the bid value which is rather important because the bidder can pay the amount of her successful bid $ b $ only when her wealth is greater than or equal to her bid i.e. $ w \geq b $. Indeed, the utility function \eqref{ufw} will yield a positive utility even when $ b >w $ (as long as $w+v \geq b$) and therefore the bidder does not have the ability to buy the item. This is unrealistic.\\
For the sake of simplicity of the notation, we assume that the auction has no entry cost. Alternatively, we could also assume that $w$ is the wealth after discounting the entry cost. Then, using \eqref{ufw}, Brenda's expected utility would be of the form 
\begin{equation} \label{pi2}
\varPsi_1=(w+v-b)^{r}F(b)+w^r[1-F(b)], 
\end{equation}
where $ F(b) $ is Brenda's probability of winning the auctioned item. The expected utility function \eqref{pi2} has been used by for example, \cite{gentry2015existence} and \cite{li2017hidden} in game theoretic perspective among others. In \eqref{pi2}, $ w^r $ has been taken as Brenda's utility in case of not winning the auctioned item with probability $ [1-F(b)] $. However, this implies that where Brenda is a risk averse  bidder $ (0<r<1) $ or a risk-seeking bidder $ (1<r\leq 2) ,$ her wealth will change when she fails to win the item. This is unrealistic too since in a typical FPSB auction, the bidder's wealth remains unchanged upon losing the bid.\\
Having $w$ unconstrained with respect to $b$ can lead to unrealistic solutions. For example, not only can the optimal bid value $b^\ast$ be greater than $w$, but also, $b^\ast$ could decrease as $w$ increases which is inconsistent with the item being normal. We illustrate this using a simple example.\\

\begin{example} \label{ex0}
Suppose that Brenda has true value $ v=\$150 $ for the auctioned item which has no reserve price. From her subjective belief about Charles, lets assume that she elicits the distribution $ F(c)=\frac{9c}{8 \times 200}-\frac{c^9}{8 \times 200^9} $ on Charles's bid  $ c $ \cite[]{banks2015adversarial}. To find her optimal bid, she can replace $ c $ by $ b $ to obtain $ F(b),$ her probability of winning the item.
Using the expected utility function \eqref{pi2}, Brenda can find her optimal bid by solving the following equation
\begin{align} \label{euf6}
b^\ast=\text{argmax}_{b \in \rm I\!R^+}[\{w+v-b\}^{r}F(b)+w^r\{1-F(b)\}].
\end{align}

\begin{table}[h!]
	\small
	\begin{center}
		\caption{Brenda's optimal bids $ b^\ast $ using  expected utility function \eqref{pi2} with different wealth and risk-aversion levels}
		\label{t3}
		\begin{tabular}{ |p{1.5cm}||p{1.8cm}|p{1.8cm}|p{1.8cm}|p{1.8cm}|}
			\hline 
			$r$&0.90&0.50&0.10&0.05\\
			\hline
			\hline 
			$ w=0 $& 78.93 &99.88 &135.84 &148.38 \\
			$ w=50 $& 76.44 &82.14&87.51 &88.15 \\
			$ w=150 $& 75.69& 78.46& 81.12 & 81.44\\
			\hline
		\end{tabular}
	\end{center}
\end{table}
Table \ref{t3} shows that $ b^\ast $ increases with increase in her risk-aversion, which is realistic. However, it also shows that $ b^\ast $ decreases with her wealth $ w $, which is unrealistic when the item is normal, because the demand of the normal item increases with wealth \cite[]{baisa2017auction}. Further, it shows that $ b^\ast $ can be much higher than her wealth $ w $ i.e. $ b^\ast >w $ especially for higher risk-aversion levels ($ r=0.10 $ or $ 0.05 $).
\end{example}

Now, if we assume that $ w=0 $, then the expected utility function \eqref{pi2} would be of the form
\begin{equation} \label{pi}
\varPsi_1=(v-b)^{r}F(b).
\end{equation}
 This function has been used for risk-averse bidders by \cite{cox1982theory} and \cite{cox1985experimental}, among others, and by \cite{banks2015adversarial} for risk-neutral bidders. The first row of Table \ref{t3} shows Brenda's optimal bids for this special case where $ w=0 $. Finding $ b^\ast $ using \eqref{pi} is also unrealistic again because $ b^\ast > w.$
 
 \subsection{New CRRA Parameters} \label{crra}
When an item is normal, the bidders' willingness to pay for it increases with their wealth \cite[]{baisa2017auction}. We propose to introduce an additional risk behaviour parameter $a$ that will change with the relative change in circumstances of the bidder's wealth. The original risk behaviour parameter $r$ will remain unchanged and will denote the baseline risk behaviour level of the bidder. This relative change in circumstances could occur in two specific cases; firstly, when the bidder's wealth changes and secondly, when bidder attempts to model the risk behaviour of their opponent. We describe how the new parameter $a$ can be defined in each of these cases.\\
Firstly, to model Brenda's risk behaviour at an increase level of her wealth compared with her own lower level of wealth, we modify the CRRA parameter as follows: We define $ r_B $ to be Brenda's baseline risk behaviour parameter which represents her natural risk appetite at her wealth, say $ w_1 $. Note that $ r_B $ is the same as the $ r $ we defined earlier in Subsection \ref{UF}. 
Lets assume that her circumstances change (e.g. she gains an inheritance) and her wealth is increased to $ w_2 $ $ (w_2>w_1) $. At this increased wealth level, we expect her to be more risk-averse (or less risk-seeking) for the same auctioned item (since the item is assumed normal). So, we modify her risk behaviour parameter having wealth $ w_2 $ relative to wealth $ w_1 $ as
\begin{equation} \label{aB}
a_B = \begin{cases*} r_B^{\frac{1}{h}} \qquad \text{if} \quad 0 <r_B<1\\
r_B \qquad \text{if} \quad r_B=1\\
r_B^h \qquad \text{if} \quad 1 <r_B \leq 2,
\end{cases*}
\end{equation}
where, we define $ 0<h= w_1/w_2<1$, when $w_1 < w_2.$ Note that here, for $ 0<r_B\leq 2 $  $ (r_B \neq 1) $, $ a_B<r_B $ i.e. if Brenda was risk-averse (risk-seeking) at wealth level $ w_1 $, then she is even more risk-averse (less risk-seeking) at wealth level $ w_2 $. Also, when $r_B=1,\,a_B=r_B $ i.e. if Brenda was risk-neutral at wealth level $ w_1 $, she is also risk-neutral at wealth level $ w_2 $. There is no change to her risk behaviour (that is, $ a_B=r_B $ ) if her wealth decreases. 

Secondly, when Brenda is bidding against Charles who has wealth $ w_C $, we modify their CRRA parameters as follows: If Brenda believes that Charles has wealth $ w_C $, we define  $ R_C $ as Charles's natural risk appetite for the auctioned item that Brenda believes.  In this case, Brenda could draw $ R_C $ from a uniform distribution $ U $ with support $ (0,1) $ if she believes that Charles is a risk-averse bidder. She could draw $ R_C $ from a uniform distribution $ U $ with support $ (1,2] $ if she believes that Charles is a risk-seeking bidder. In this case, we assume that Brenda has wealth $ w_B $ where $ w_B>w_C $ i.e. Brenda has more wealth than Charles. Thus, Brenda's risk behaviour parameter in this case would be same as defined in \eqref{aB} with $ 0<h= w_C/w_B<1$.
However, if Brenda believes that Charles has more wealth than her i.e. $ w_C >w_B $, she can modify Charles's risk-behaviour parameter as
\begin{equation} \label{AC}
A_C = \begin{cases*} R_C^{\frac{1}{h}} \qquad \text{if} \quad 0 <R_C<1\\
R_C \qquad \text{if} \quad R_C=1\\
R_C^{h} \qquad \text{if} \quad 1 <R_C \leq 2,
\end{cases*}
\end{equation}
where $ 0<h=w_B/w_C<1 $ and $ A_C <R_C $ if Brenda believes that Charles is risk-averse (risk-seeking) bidder in this case. Thus, $ A_C $ may take values in the interval $(0,1) $ if she believes that he is a risk-averse bidder. On the other hand if she believes that he is a risk-seeking bidder, $ A_C $ may take values in the interval $ (1,2^{h}] $ and $ A_C=1 $, if she believes that he is a risk-neutral bidder. In this case, Brenda's risk behaviour parameter would remain unchanged i.e. it is $ a_B=r_B $ for $ 0<r_B \leq 2 $. 

\subsection{New CRRA Utility Function} \label{uf1}
We propose to modify the utility function so that $ w \geq b $ i.e. a bid value can not be greater than wealth. Additionally, since in equilibrium, bidders never bid above their true values \cite[]{gentry2015existence}, we in fact,  have that $ b \leq v \leq w$, because she can bid and pay an amount $ b \leq v $ only if her wealth is at least equal to her true value. Further, we assume that a person's wealth remains unchanged if she loses her bid.\\
We propose to use the following utility function for Brenda
\begin{align} \label{puf}
u(b,v,w)=\begin{cases*} w+(v-b)^{a_{B}} \qquad  &\text{if she wins the bid}\\
w \qquad  &\text{if she loses the bid},
\end{cases*}
\end{align}
where, $ a_{B} $ is a modified CRRA parameter defined in Section \ref{crra}. 
Thus, for our proposed utility function \eqref{puf}, Brenda's expected utility would be of the form
\begin{equation} \label{eub}
\varPsi_2=\{w+(v-b)^{a_B}\}F(b)+w\{1-F(b)\}.
\end{equation}
The above equation simplifies to
\begin{equation} \label{eubs}
\varPsi_2=w+(v-b)^{a_B}F(b).
\end{equation}
Thus, using \eqref{eubs}, Brenda can find her optimal bid by solving the following equation
\begin{align} \label{euf5}
b^\ast=\text{argmax}_{b \in \rm I\!R^+}[w+(v-b)^{a_B}F(b)].
\end{align}
The utility function \eqref{puf} is more realistic in the sense that it allows Brenda to add up her profit to her wealth after the successful bid. It also allows Brenda to bid strictly less than or equal to her true value $ v $ and consequently to bid less than or equal to her wealth at any assumed level of her risk-aversion. Lets assume that Brenda has wealth $ w=\$150 $, true value $ v=\$150 $ and $ F(b) $ as considered earlier in this Section. Then, using Equation \eqref{euf5}, she could get her optimal bids as shown in first row of Table \ref{t3} for different assumed risk-aversion levels. Note that she will get these values since her optimal bid is not affected by her wealth in Equation \eqref{euf5}. Thus, this function gives the leverage to assume any wealth $ w \geq v $ for Brenda and also her optimal bid to be bounded above by her wealth for any assumed risk-aversion level i.e. $ b^\ast \leq v \leq w $.

We define \eqref{puf} as Brenda's wealth plus her profit where she could be risk-neutral, risk-averse or risk-seeking in her profit. Thus, by letting her profit $ v-b=x $, we can show that \eqref{puf} is a CRRA utility function since
\begin{align*}
\frac{d}{dx}[u(x,w)]=&u'(x,w)=a_{B}x^{a_{B}-1}\\
\frac{d}{dx}[u'(x,w)]=&u''(x,w)=a_{B}(a_{B}-1)x^{a_{B}-2}\\
\text{so}\\
\implies \frac{u''(x,w)}{u'(x,w)}=&\frac{a_{B}(a_{B}-1)x^{a_{B}-2}}{a_{B}x^{a_{B}-1}}\\
\implies 1-a_{B}=&-x\frac{u''(x,w)}{u'(x,w)},
\end{align*}
where $ (1-a_{B}) $ is the coefficient of CRRA that incorporates the effect of increase in wealth on bidder's risk behaviour and is defined in Section \ref{crra}.

\section{Non-Strategic Play} \label{NSP}
In this section, we show how an ARA solution for Brenda's optimal bid can be found when she assumes that Charles is a non-strategic opponent. In this case, Charles will bid an amount that is independent of Brenda's bid. 
We assume that Brenda bids an amount $ b $, having wealth $ w_B $ and true value $ v_B $ for the auctioned item. We assume that the auctioned item is normal and it has a reserve price $ \tau $ such that $ \tau <b \leq v_B \leq w_B $. We assume that $ \tau $ is known in advance to each bidder. We define Brenda's wealth $ w_B $, as the money she has at her disposal. She does not know about Charles's wealth $ W_C $ and places a distribution $ H_{BC} $ on his wealth. She also does not know about Charles's true value $ V_C $ and his bid $ C $ for the auctioned item. So, she places a distribution $ G_{BC} $ on his true value according to her belief and then finds the distribution of his bid $ C $ as defined in Equation \eqref{dfi} below.  For Charles, she also believes that $ \tau <c \leq v_C \leq w_C $ holds, where $ v_C $ and $ w_C $ are chosen by Brenda from the distributions $ G_{BC} $ and $ H_{BC} $ respectively.

Brenda's probability of winning from a bid of amount $ b $ is given by
\[
F_{BC}(b) = p(C\leq b),
\]	
where, $ F_{BC} $ is the distribution over Charles's bid with support $ (\gamma, \kappa] \subseteq  \rm I\!R^+$ with $ 
\gamma \geq \tau $ that Brenda believes and $C$ is the random variable of Charles's bid that she believes. To obtain $F_{BC}$, Brenda divides her introspection into two parts as $G_{BC}$, the cumulative distribution function (CDF)  that quantifies her uncertainty of Charles's true value and $T_{BC}$, the CDF that quantifies her uncertainty for the fraction of Charles's true value $p = c/v_C $ that he bids. Note that the supports for $G_{BC}$  and $T_{BC}$ are $ (\gamma, \kappa] $ and $ (\gamma/v_C, 1],$ respectively.\\
She can then derive her subjective distribution function over $ C=PV_C $, the amount of Charles's bid as
\begin{align*} 
F_{BC}(c)={\rm I\!P}[\gamma < PV_C\leq c]
& =\int_{\gamma }^{c}\int_{\gamma/v_C}^{1}g_{BC}(v_C)t_{BC}(p)\,dp\,dv_C\\
&+\int_{c}^{\kappa}\int_{\gamma/v_C}^{c/v_C}g_{BC}(v_C)t_{BC}(p)\,dp\,dv_C. \numberthis \label{dfi1}
\end{align*}

\begin{figure}[h!]
	\begin{centering}
		\scalebox{0.60}{\input{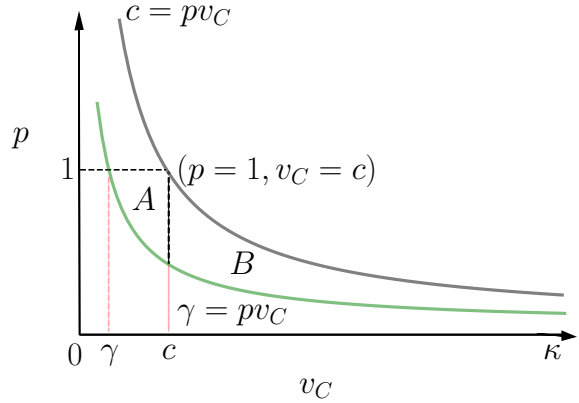}}
		\caption{Region of integration when the bid is a proportion of the true value with reserve price $ \tau $.}
		\label{dist plot 3}
	\end{centering}
\end{figure}

As $ \int_{\gamma/v_C}^{1}t_{BC}(p)dp = 1 $, the above equation simplifies to
\begin{equation} \label{dfi}
\begin{split}
F_{BC}(c)& = G_{BC}(c)+\int_{c}^{\kappa}g_{BC}(v_C)T_{BC}(c/v_C)\,dv_C,
\end{split}	
\end{equation}
where, $ g_{BC}(v_C) $ is the probability density function for Charles's true value that Brenda elicits and $ t_{BC}(p) $ is the probability density function for the fraction of Charles's true value that Brenda believes he will bid. Equation (\ref{dfi1}) assumes that Charles's true value $ V_C $ and fraction of his true value $ P $ are independent. In Equation (\ref{dfi1}), the whole region of integration has been divided into two regions for the random variables $ V_C $ and $ P $ in order to find the distribution of $ C $. Figure \ref{dist plot 3} shows the division of the integration region into these two regions. Area between the two curves shows the integration region between $ \gamma $ and $ \kappa $. The area $ A $ corresponds to first integral part whereas area $ B $ corresponds to second integral part of Equation \eqref{dfi1}.\\

Then, using \eqref{puf}, we rewrite Brenda's expected utility function \eqref{eub} as
\begin{equation} \label{euf} \nonumber
\begin{split}
\varPsi_B=&\{w_B+(v_B-b)^{a_B}\}F_{BC}(b)+w_B\{1-F_{BC}(b)\}.
\end{split}	
\end{equation}
This Equation simplifies to
\begin{equation} \label{eufs} 
\begin{split}
\varPsi_B=&w_B+(v_B-b)^{a_B}F_{BC}(b),
\end{split}	
\end{equation}
where $ F_{BC}(b) $ is her probability that her bid $ b $ will be successful. Finally, Brenda's optimal bid $ b^\ast $ may be found by solving the following equation
\begin{align} \label{nsp}
b^\ast=\text{argmax}_{b > \gamma}[w_B+(v_B-b)^{a_B}F_{BC}(b)].
\end{align}
Numerical methods may often be needed to solve \eqref{nsp} for $ b^\ast $. Note that $b^\ast$ is, in fact, a function of $a_{B}$ and therefore, a function of $r_B$ and $w_C.$ Of which, $r_B,$ being the baseline risk behaviour, is constant for a given player. Therefore, we can re-write \eqref{nsp} as 
\begin{align} \label{nsp1}
b^\ast (w_C) =\text{argmax}_{b > \gamma}[w_B+(v_B-b)^{a_B}F_{BC}(b)].
\end{align}
If Brenda has information on Charles's wealth, she can use \eqref{nsp1} to find her optimal bid amount. Alternatively, she can take into consideration her uncertainty around $w_C$ and find the expected value of her optimal bid as
\begin{align} \label{nsp2}
\mathop{\mathbb{E}}(b^\ast)=\int b^\ast(w_C) \, dH_{BC}(w_C).
\end{align}

Comparing the derivation above with the ARA sketch provided in Section \ref{ara} the reader can note that $F_{BC}(b)$ of \eqref{dfi} is the $p_B(c)$ in \eqref{eu4} obtained by assuming that the opponent is a non-strategic player and that \eqref{eufs} provides the expected utility $\varPsi_B(b)$ defined in \eqref{eu2} for this particular problem. Also, note that in the above analysis, we have assumed that all the probability distributions considered are continuous. If any of the distributions are discrete then the corresponding integrals would be replaced by summations.\\

\begin{example}	\label{ex1}	
Suppose Brenda's true value for the item $ v_B=\$150 $, her wealth $ w_B=\$150 $ and the auctioned item has a reserve price $ \tau $. She could elicit her uncertainty on Charles's true value using a uniform distribution $ G_{BC}(v_C)=\frac{(v_C-\gamma)}{\kappa-\gamma} $ with support $ (\gamma, \kappa], \, \gamma \geq \tau$. Since she believes that Charles is a non-strategic player whose bid amount will be independent of Brenda's bid, she could assume a distribution\footnote{The distributions $ T_{BC}(p)$ and $H_{BC}$ elicited here are taken from \cite{banks2015adversarial} but with reserve price $ \tau $ that constrains $\gamma$.} on $p,$ the proportion of Charles's true value that Brenda believes he will bid, given by  $ T_{BC}(p)=\frac{p^8-(\gamma/v_C)^8}{1-(\gamma/v_C)^8} $ with support $ (\gamma/v_C < p \leq 1] $. Finally, suppose that she has a uniform distribution $ H_{BC}(w_C)=\frac{w_C-\alpha}{\beta-\alpha} $ on Charles's wealth $ W_C $ with support $ (\alpha, \beta]$ where $\alpha \geq \gamma.$
Then we have that
	\[
	g_{BC}(v_C) = \frac{1}{\kappa-\gamma}.
	\]
	Using Equation \eqref{dfi} where $ c $ is replaced by $ b $, we get
	\footnote{The computations in this paper have been performed by using Maple™}
\begin{align*}
	F_{BC}(b) =& \frac{b-\gamma}{\kappa-\gamma}+\frac{b^8-\gamma^8}{\kappa-\gamma}\bigg[-\frac{\sqrt{2}}{16 \times \gamma^7} \textnormal{ln}\Big(\frac{\kappa^2+\kappa \gamma \sqrt{2}+\gamma^2}{\kappa^2-\kappa \gamma \sqrt{2}+\gamma^2}\Big)\\
	&+\frac{\sqrt{2}}{16 \times \gamma^7} \textnormal{ln}\Big(\frac{b^2+b \gamma \sqrt{2}+\gamma^2}{b^2-b \gamma \sqrt{2}+\gamma^2}\Big)
	-\frac{\sqrt{2}}{8 \times \gamma^7} \textnormal{tan$^{-1}$}\Big(\frac{\kappa \sqrt{2}}{\gamma}+1\Big)\\
	&+\frac{\sqrt{2}}{8 \times \gamma^7} \textnormal{tan$^{-1}$}\Big(\frac{b \sqrt{2}}{\gamma}+1\Big)-\frac{\sqrt{2}}{8 \times \gamma^7} \textnormal{tan$^{-1}$}\Big(\frac{\kappa \sqrt{2}}{\gamma}-1\Big)\\
	&+\frac{\sqrt{2}}{8 \times \gamma^7} \textnormal{tan$^{-1}$}\Big(\frac{b \sqrt{2}}{\gamma}-1\Big)-\frac{1}{4 \times \gamma^7}\textnormal{tan$^{-1}$}\Big(\frac{\kappa}{\gamma}\Big) +\frac{1}{4 \times \gamma^7}\textnormal{tan$^{-1}$}\Big(\frac{b}{\gamma}\Big)\\
	&+\frac{\textnormal{ln}(\kappa-\gamma)}{8 \times \gamma^7}-\frac{\textnormal{ln}(b-\gamma)}{8 \times \gamma^7}-\frac{\textnormal{ln}(\kappa+\gamma)}{8 \times \gamma^7} +\frac{\textnormal{ln}(b+\gamma)}{8 \times \gamma^7} \bigg] \numberthis \label{cdf}. 
\end{align*}
Suppose $ \tau=\$30 $ and Brenda believes that $ G_{BC} $ has support $ (30, 200] $. Then substituting these in \eqref{cdf}, she can get $ F_{BC} (b)$ which she can then use to find her optimal bid by solving Equation \eqref{nsp} which also takes into consideration her risk appetite. 
	
Now, assuming that Brenda is a risk-neutral bidder i.e. $ a_B=1 $, her optimal bid by solving Equation \eqref{nsp} for $ w_B=v_B=\$150 $ turns out to be $\$88.05$ with probability of winning of $0.415$ and with the expected utility of $175.72$. 
	
Next, we assume that Brenda is a risk-averse bidder. For the sake of simplicity, we assume that Brenda chooses Charles's wealth $ w_C$ to be $\$150 $ from the distribution $ H_{BC} $. We assume that Brenda's baseline risk behaviour parameter is $ r_B $ when her wealth $ w_B=\$150 $. She believes that Charles's baseline risk behaviour parameter is $ R_C $ at $ w_C=\$150 $. If Brenda's wealth was to increase to $ w_B=\$200 $, then, we expect her to be more risk-averse than Charles because she is able to (and willing to since the item is normal) pay a little more to increase her chance of winning the bid. By using \eqref{aB}, we model Brenda's risk-aversion when $ w_B=\$200 $ relative to $ w_B=\$150 $ as $ a_B=r_B^{\frac{1}{h}}=r_B^ {\frac{200}{150}}=r_B^{1.33} $. In Table \ref{NSPRA}, we show how Brenda's optimal bids, her probabilities of winning and her expected utilities change with change in her wealth and also how they change with change in $ r_B $. It shows that with increase in her wealth to $ w_B=\$200 $, she is more risk-averse and consequently bids higher than when $ w_B=\$150 $. In general, it shows that an increase in risk aversion leads to higher optimum bid (resulting in a higher probability of winning that bid) but a lower expected utility nonetheless. We also plot these in Figure \ref{nspra}.
	
	\begin{table}[h!] 
		\small
		\begin{center}
			\caption{Brenda's optimal bids, probabilities of winning and her expected utilities when she is risk-averse and risk-neutral bidder and she thinks that $w_C = \$150.$} \label{NSPRA}
			\begin{tabular}{ |p{2.2cm}||p{.88cm}|p{.88cm}|p{.88cm}|p{.88cm}|p{.88cm}|p{.88cm}|p{.88cm}|p{.88cm}|p{.88cm}|p{.88cm}|}
				\hline 
				$a_B=r_B$ $ (w_B=\$150) $&0.10&0.20&0.30&0.40&0.50&0.60&0.70&0.80&0.90&1.00\\
				\hline
				$ b^\ast(w_C) $& 138.06 &128.69 &120.89 &114.22 &108.43&103.36&98.87&94.86&91.28&88.05\\
				$ F_{BC}(b^\ast) $& 0.743 &0.684 &0.633 &0.589 &0.551&0.518&0.488&0.461&0.437&0.415\\
				$ \varPsi_B $& 150.95 &151.26&151.74 &152.46 &153.55&155.19&157.66&161.40&167.08&175.72 \\
				\hline
				\hline
				
				$ a_B=r_B^{1.33} $ $(w_B=\$200)$ &0.047&0.118&0.202&0.296&0.398&0.507&0.622&0.743&0.869&1.00\\
				\hline	
				$ b^\ast(w_C)$& 143.95 &136.22 &128.52 &121.17 &114.34&108.05&102.32&97.09&92.35&88.05\\
				$ F_{BC}(b^\ast) $& 0.779 &0.732 &0.683 &0.635 &0.590&0.549&0.511&0.476&0.444&0.415\\
				$ \varPsi_B $& 200.85 &201.00&201.27 &201.72 &202.45&203.65&205.65&209.08&215.05&225.72 \\
				\hline	
			\end{tabular}
		\end{center}
	\end{table}

But often, Brenda may instead be uncertain about $W_{C}$ and may prefer to take into account her uncertainty and obtain her expected optimal value after doing so. To do this, she can first elicit $H_{BC}$ and then solve \eqref{nsp2} using Monte Carlo methods. Suppose she believes that Charles's wealth is uniformly distributed between $\$100$ and $\$300.$ She will derive $a_{B}$ as described in Section \ref{crra}, whereby, $a_{B} = r_{B}^{1/h},$ for $h=w_{C}/w_{B},$ when $w_{B}>w_{C}$ and $a_{B} = r_{B}$ otherwise. Since $W_{C}$ is a random variable, so will be $A_{B}.$ Then, her expected optimal bids, probabilities of winning and expected utilities thus obtained are given in Table \ref{NSPRAD}.
\begin{table}[h!] 
	\small
	\begin{center}
		\caption{Brenda's optimal bids, probabilities of winning and her expected utilities when she is risk-averse and risk-neutral bidder and draw $ W_C $ from $ H_{BC} $} \label{NSPRAD}
		\begin{tabular}{ |p{2.65cm}||p{.88cm}|p{.88cm}|p{.88cm}|p{.88cm}|p{.88cm}|p{.88cm}|p{.88cm}|p{.88cm}|p{.88cm}|p{.88cm}|}
			\hline 
			$r_B$ $(w_B=\$200)$ &0.10&0.20&0.30&0.40&0.50&0.60&0.70&0.80&0.90&1.00\\
			\hline 
			\hline
			$\mathop{\mathbb{E}}(a_B)$ &0.07&0.16&0.25&0.35&0.44&0.55&0.66&0.77&0.88&1.00\\
			$ \mathop{\mathbb{E}}(b^\ast) $&140.91&132.52&124.73&117.94&111.84&106.19&100.89&96.21&91.89&88.05\\
			$ F_{BC}[\mathop{\mathbb{E}}(b^\ast)] $&0.761&0.708&0.658&0.614&0.574&0.536&0.501&0.470&0.441&0.415\\
			$ \varPsi_B[\mathop{\mathbb{E}}(b^\ast)] $&200.89&201.12&201.48&202.07&202.85&204.29&206.55&210.11&215.74&225.72\\
			\hline	
		\end{tabular}
	\end{center}
\end{table}
	
	\begin{figure}[h]
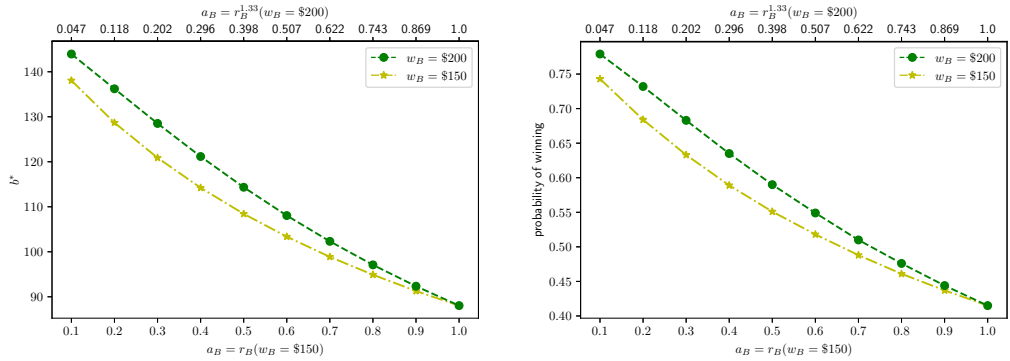
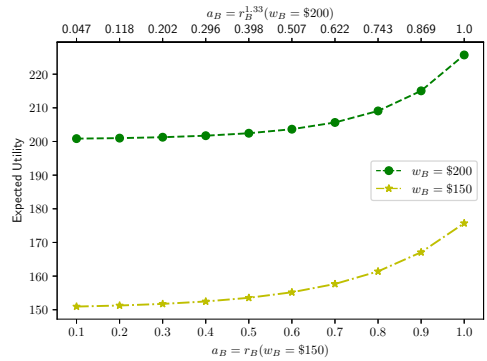

		\scriptsize
		\centering
		\begin{subfigure}[t]{0.48\textwidth}
			\centering
			\scalebox{0.48}{\input{nspra1.pgf}}
			\caption{Brenda's optimal bids}
			\label{nspra1}
		\end{subfigure}
		\hfill
		\begin{subfigure}[t]{0.48\textwidth}
			\centering
			\scalebox{0.48}{\input{nspraF.pgf}}
			\caption{Brenda's probabilities of winning}
			\label{nspraF}
		\end{subfigure}
		\hfill

		\begin{subfigure}[t]{0.48\textwidth}
			\centering
			\scalebox{0.48}{\input{nsprau1.pgf}}
			\caption{Brenda's expected utilities}
			\label{nsprau1}
		\end{subfigure}		
		\caption{Comparison of Brenda's (risk-averse and risk-neutral) optimal bids, her probabilities of winning and her expected utilities when she has $ w_B=\$150 $ and $ \$200 $}
		\label{nspra}
	\end{figure}

Finally, we find Brenda's optimal bids, probabilities of winning and expected utilities when she is assumed to be a risk-seeking bidder and $w_{B} =\$ 150$. These are listed in Table \ref{NSPRS}. These show that as the risk seeking behaviour intensifies, the optimal bids get lower and so does the probabilities of winning the bids. Again, we can model how her bids will change if her wealth was to increase, to say $\$200.$ By using \eqref{aB}, we model Brenda's risk-seeking behaviour when $ w_B=\$200 $ as $ a_B=r_B^h=r_B^ {{150}/{200}}=r_B^ {0.75} $. 
It shows that with increase in her wealth i.e. $ w_B=\$200 $, she is less risk-seeking and consequently she bids higher than when $ w_B=\$150 $. In general, it shows that an increase in risk seeking behaviour leads to a lower optimum bid (resulting in a lower probability of winning that bid) but a higher expected utility nonetheless.
We also plot these results in Figure \ref{nsprs}.

	\begin{table}[H] 
		\small
		\begin{center}
			\caption{Brenda's optimal bids, probabilities of winning and her expected utilities when she is risk-seeking and risk-neutral bidder and she thinks that $w_C = \$150.$} 
			\label{NSPRS}
			\begin{tabular}{ |p{2.2cm}||p{.88cm}|p{.88cm}|p{.88cm}|p{.88cm}|p{.88cm}|p{.88cm}|p{.88cm}|p{.88cm}|p{.88cm}|p{1.1cm}|}
				\hline
				$a_B=r_B$ $ (w_B=\$150) $&1.00&1.10&1.20&1.30&1.40&1.50&1.60&1.70&1.80&1.90\\
				\hline				
				$ b^\ast(w_C)$&88.05& 85.12&82.45&80.02&77.79&75.73 &73.84&72.08&70.45&68.93\\				
				$ F_{BC}(b^\ast)$&0.415& 0.400&0.378&0.361&0.346&0.332 &0.320&0.308&0.297&0.287\\				
				$ \varPsi_B $&175.72& 189.39 &209.30&240.36 &288.39 &362.50&478.03&656.24&933.26&1365.39 \\
				\hline
				\hline
				
				$a_B=r_B^{0.75 }$ $ (w_B=\$200) $&1.000 &1.074&1.147&1.217&1.287&1.355&1.423&1.489&1.554&1.618\\				
				\hline
				$ b^\ast(w_C)$&88.05&85.85 &83.84 &82.02&80.32 &78.77&77.30&75.95&74.69&73.51\\				
				$ F_{BC}(b^\ast)$&0.415&0.400 &0.387 &0.375& 0.363&0.353&0.343&0.334&0.326&0.318\\				
				$ \varPsi_B $&225.72& 234.91&247.42&263.68 &285.50 &314.32&352.85&402.99&469.06&554.90 \\				
				\hline
			\end{tabular}
		\end{center}
	\end{table}

	\begin{figure}[h]
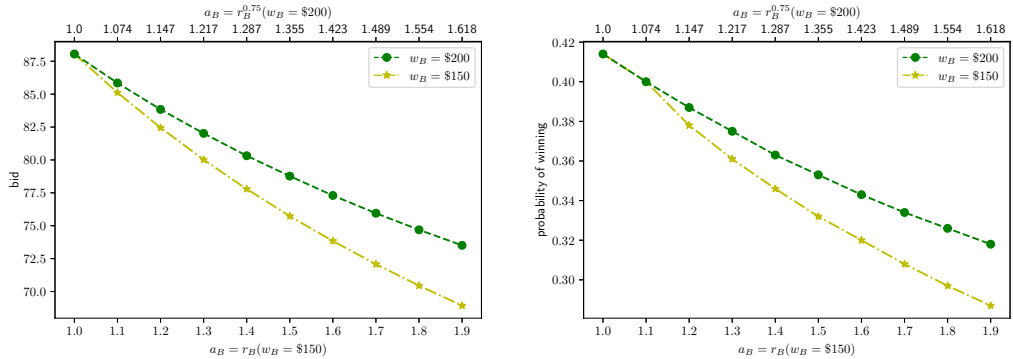
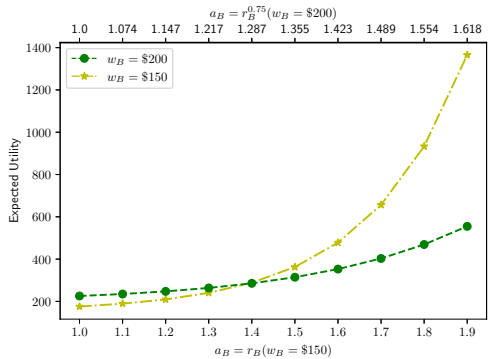

		\scriptsize
		\centering
		\begin{subfigure}[t]{0.48\textwidth}
			\centering
			\scalebox{0.48}{\input{nsprs1.pgf}}
			\caption{Brenda's optimal bids}
			\label{nsprs1}
		\end{subfigure}
		\hfill
		\begin{subfigure}[t]{0.48\textwidth}
			\centering
			\scalebox{0.48}{\input{nsprsF1.pgf}}
			\caption{Brenda's probabilities of winning}
			\label{nsprsF1}
		\end{subfigure}
		\hfill

		\begin{subfigure}[t]{0.48\textwidth}
			\centering
			\scalebox{0.48}{\input{nsprsu1.pgf}}
			\caption{Brenda's expected utilities}
			\label{nsprsu1}
		\end{subfigure}
		\caption{Comparison of Brenda's (risk-seeking and risk-neutral) optimal bids, her probabilities of winning and her expected utilities when she has $ w_B=\$150 $ and $ \$200 $}
		\label{nsprs}
	\end{figure}
 
 Once again, Brenda may prefer to take into account her uncertainty in $W_{C}$ and obtain her expected optimal value after doing so. To do this, she can solve \eqref{nsp2} using Monte Carlo methods. Her expected optimal bids, probabilities of winning and expected utilities thus obtained are given in Table \ref{NSPRSD}.

\begin{table}[h!] 
	\small
	\begin{center}
		\caption{Brenda's optimal bids, probabilities of winning and her expected utilities when she is risk-seeking and risk-neutral bidder and draw $ W_C $ from $ H_{BC} $} \label{NSPRSD}
		\begin{tabular}{ |p{2.65cm}||p{.88cm}|p{.88cm}|p{.88cm}|p{.88cm}|p{.88cm}|p{.88cm}|p{.88cm}|p{.88cm}|p{.88cm}|p{1cm}|}
			\hline 
			$r_B$ $ (w_B=\$200) $&1.00&1.10&1.20&1.30&1.40&1.50&1.60&1.70&1.80&1.90\\
			\hline
			\hline
			$\mathop{\mathbb{E}}(a_B)$
			&1.00&1.09&1.17&1.26&1.34&1.42&1.51&1.60&1.68&1.76\\
			$ \mathop{\mathbb{E}}(b^\ast) $&88.05&85.49&83.18&81.01&79.11&77.39&75.62&74.00&72.54&71.30\\
			$ F_{BC}[\mathop{\mathbb{E}}(b^\ast)] $&0.415&0.398&0.383&0.368&0.355&0.344&0.332&0.320&0.311&0.301\\
			$ \varPsi_B[\mathop{\mathbb{E}}(b^\ast)] $&225.72&237.36&252.22&276.33&307.23&350.94&422.22&527.87&664.06&857.73\\
			\hline	
		\end{tabular}
	\end{center}
\end{table}

\end{example}
\section{Level-\textit{k} Thinking} \label{k}
We will now show how an ARA solution for a two player FPSB auction game can be derived when Brenda believes that Charles is a level-\textit{k} thinker. In a level-\textit{k} analysis, we model how deeply does the opponent think about the problem \cite[][]{stahl1995players'}. It is not only an important solution concept to be considered for a strategic adversary \cite[]{banks2015adversarial} but also provides the flexibility to model the opponent at different levels of strategic thinking including being a non-strategic player (when $k=0$). If the decision maker performs a level-\textit{k} analysis, she believes that her opponent is a level-$(k-1)$ thinker who would model her as a level-$(k-2)$ player. For instance, when $k=1,$ she believes that her opponent is a level-0 thinker, that is, a non-strategic player. A level-2 analysis means that the decision maker assumes that her opponent is a level-1 thinker, who believes that she is a level-0 thinker. A level-3 analysis means that the decision maker assumes that her opponent is a level-2 thinker who models her as a level-1 thinker and so on. So, in such type of modelling, the decision maker attempts to think one level deeper than her opponent.

The key question is how large should the \textit{k} be? While in principle, \textit{k} could take higher values, \cite{ho1998iterated} and \cite{lee2012game} present experimental evidence that people do not usually think higher than level $ 2 $ or $ 3 $. Therefore, in practice, it makes sense to solve the level-\textit{k} problem for $k$ being 1, 2 or 3.

As described earlier, we extend the work by \cite{banks2015adversarial}. We derive Brenda's optimal bid for the level-\textit{k} thinking solution concept assuming not only that she and Charles are risk-neutral bidders but also when they are risk-averse or risk-seeking bidders. Also, we consider their wealth as well as assume that the auctioned item has a reserve price $ \tau $ when deriving the optimal bid.

For \textit{k} $=1 $, where Brenda believes herself to be a level-1 thinker and models Charles as a level-0 (non-strategic) thinker, the problem is identical to the non-strategic thinking problem modeled in Section \ref{NSP}.

Here, we derive ARA solution for the case where \textit{k} $=2 $. In this case, Brenda models herself as a level-2 thinker and believes that Charles is a level-1 thinker, who (Charles) models Brenda as a level-0 thinker. Brenda performs ARA by placing a subjective distribution $ G_{BCB} $ with support $ (\gamma_1, \kappa_1] $ on her true value $ V_B $ that Charles might elicit, a distribution $T_{BCB}$ with support $ (\gamma_1/V_B, 1] $ on the fraction of her true value that she thinks Charles would think that she would bet. This allows her to derive $F_{BCB}$ using \eqref{dfi} (but with roles reversed) with support $ (\gamma_1, \kappa_1] $ that she believes is the distribution of her bid that Charles might elicit. She would then elicit a distribution $H_{BCB}$ on her wealth with support $ (\eta_1, \beta_1],$ where $ \eta_1 \geq \gamma_1 $ and $ \beta_1 \geq \kappa_1,$ that she thinks Charles would elicit, a distribution $ G_{BC} $ with support $ (\gamma_2, \kappa_2] $ on Charles's true value $ V_C $, and $ H_{BC} $ with support $ (\eta, \beta] $ on Charles's wealth $ W_C $ where $ \eta \geq \gamma_2 $ and $ \beta \geq \kappa_2 $. She can then find his optimal bid for given $w_{B}, w_{C}, v_{C}$ and $r_{C}$ as 
\begin{equation} \label{cl1ra}
C^\ast (w_{B}, w_{C}, v_{C}, r_{C})=\text{argmax}_{c > \gamma_1}[w_C+(v_C-c)^{a_C}F_{BCB}(c)],
\end{equation}
where $ \gamma_1 \geq \tau $ and $ A_C $ is defined in Equation \eqref{AC}, where, $H = W_B/W_C$. As described in Section \ref{crra}, Brenda could elicit a distribution on $R_{C}.$ This distribution, along with $H_{BCB}$ and $H_{BC}$ would allow her to derive the distribution for $A_{C}$ which we shall denote by $S_{BC}.$ She can then find the expected value of Charles's optimal bid that she thinks he will derive as a level-1 thinker as
\begin{align} \label{nsp5}
\mathop{\mathbb{E}}(C^\ast)=\int \int C^\ast(w_{B}, w_{C}, v_{C}, r_{C}) \, dG_{BC}(v_C) \, dS_{BC} (a_{c}).
\end{align}
Using $ \mathop{\mathbb{E}}(C^\ast) $ and a representative value of $V_{C} \sim G_{BC}$ (say, $\mathop{\mathbb{E}}(V_{C})$) she can find  $\frac{\mathop{\mathbb{E}}(C^\ast)}{\mathop{\mathbb{E}}(v_C)}=q $, the fraction of Charles's true value that Brenda believes he may bid. Since, 
\begin{equation} \label{tr}
C=qV_C,
\end{equation}
assuming that $ q $ is fixed, she can find the distribution $ F_{BC} $ of Charles's bid $ C $ using the change of variable formula as
\begin{equation} \label{fbc_levelk}
f_{BC}=|J| g_{BC}=\frac{g_{BC}}{q}.
\end{equation}
Finally, she would obtain her optimal bid for a given value of $w_{C}$ by solving

\begin{equation} \label{level KB}
b^\ast (w_{c})=\text{argmax}_{b > \delta}[w_B+(v_B-b)^{a_B}F_{BC}(b)],
\end{equation}
where $ \delta=q\gamma_2 $ and $ a_B $ is Brenda's risk-behaviour parameter defined in Equation \eqref{aB}. If Brenda has information on Charles's wealth, she can use \eqref{level KB} to find her optimal bid amount. Alternatively, she can take into consideration her uncertainty around $w_C$ and find the expected value of her optimal bid using \eqref{nsp2}.

Comparing the derivation above with the ARA sketch provided in Section \ref{ara}, the reader can note that $f_{BC}(b)$ of \eqref{fbc_levelk} gives the $p_B(c)$ in \eqref{eu4} obtained by assuming that the opponent is a level-1 thinker and that \eqref{eufs} provides the expected utility $\varPsi_B(b)$ defined in \eqref{eu2} for this particular problem. Also, note that in the above analysis, we have assumed that all the probability distributions considered are continuous. If any of the distributions are discrete then the corresponding integrals would be replaced by summations.  

\begin{example} \label{ex5}
    Suppose Brenda's true value for the item $ v_B=\$150 $, her wealth $ w_B=\$200 $ and the auctioned item has a reserve price $ \tau = \$30.$ She, as a level-2 thinker, thinks that Charles is a level-1 thinker who would model her as a non-strategic player (level-0 thinker). She believes that: 
    
    \begin{itemize}
	\item {Charles would model her true value for the auctioned item as being uniformly distributed with support $(\$30, \$200].$ That is, $G_{BCB}=\frac{(v_B-30)}{200-30} .$}
	\item {Charles would elicit his uncertainty around the fraction of her true value that she would bid as $T_{BCB} = \frac{p^8-(30/v_C)^8}{1-(30/v_C)^8} $ with support $ (30/v_C < p \leq 1] $.}
	\item {Charles would elicit his uncertainty around her wealth to be uniformly distributed on $(\$150, \$250].$ That is, $H_{BCB}=\frac{w_B-150}{250-150} .$}
	\item {Charles's true value is uniformly distributed on $(\$100, \$200].$ That is, $G_{BC}=\frac{(v_C-100)}{200-100}.$} 
	\item {Charles's wealth is uniformly distributed on $(\$100, \$300].$ That is, $ H_{BC}=\frac{(w_C-100)}{300-100}.$}
\end{itemize}

Once she has elicited these, then following the derivation above and using Monte Carlo approximations, she can derive her optimal expected bid $ \mathop{\mathbb{E}}(b^\ast),$ her probability of winning that optimal bid $ F_{BC}[\mathop{\mathbb{E}}(b^\ast)]$ and her expected utility at that optimal bid $ \varPsi_B[\mathop{\mathbb{E}}(b^\ast)]$ for various levels of her risk behaviour as well as for the various levels of risk behaviour of Charles. These are summarised in the Tables \ref{LKBRAD} to \ref{LKRARSD} below.

\begin{table}[H] 
	\small
	\begin{center}
		\caption{Brenda's optimal bids, probabilities of winning and her expected utilities when she is a risk-averse bidder and she assumes that Charles is also a risk-averse bidder} \label{LKBRAD}
		\begin{tabular}{ |p{2.65cm}||p{.88cm}|p{.88cm}|p{.88cm}|p{.88cm}|p{.88cm}|p{.88cm}|p{.88cm}|p{.88cm}|p{.88cm}|p{.88cm}|}
			\hline 
			$r_B$ $ $&0.10&0.20&0.30&0.40&0.50&0.60&0.70&0.80&0.90&1.00\\
			\hline 
			\hline
			$\mathop{\mathbb{E}}(a_B)$ &0.07&0.16&0.25&0.35&0.44&0.55&0.66&0.77&0.88&1.00\\
			$ \mathop{\mathbb{E}}(b^\ast) $&144.88&139.76&135.08&130.85&127.05&123.56&120.31&117.42&114.83&112.48\\
			$ F_{BC}[\mathop{\mathbb{E}}(b^\ast)] $&0.933&0.864&0.802&0.746&0.695&0.648&0.605&0.566&0.532&0.500\\
			$ \varPsi_B[\mathop{\mathbb{E}}(b^\ast)] $&201.05&201.25&201.58&202.10&202.76&203.52&205.67&208.28&212.20&218.78\\
			\hline	
		\end{tabular}
	\end{center}
\end{table}

\begin{table}[H] 
	\small
	\begin{center}
		\caption{Brenda's optimal bids, probabilities of winning and her expected utilities when she is a risk-seeking bidder and she assumes that Charles is a risk-averse bidder} \label{LKRSRAD}
		\begin{tabular}{ |p{2.65cm}||p{.88cm}|p{.88cm}|p{.88cm}|p{.88cm}|p{.88cm}|p{.88cm}|p{.88cm}|p{.88cm}|p{.88cm}|p{.88cm}|}
			\hline 
			$r_B$ $ $&1.00&1.10&1.20&1.30&1.40&1.50&1.60&1.70&1.80&1.90\\
			\hline 
			\hline
			{$\mathop{\mathbb{E}}(a_B)$} &1.00&1.09&1.17&1.26&1.34&1.43&1.52&1.60&1.68&1.76\\
			$ \mathop{\mathbb{E}}(b^\ast) $&112.48&110.92&109.50&108.18&107.01&105.91&104.83&103.88&103.03&102.22\\
			$ F_{BC}[\mathop{\mathbb{E}}(b^\ast)] $&0.500&0.480&0.461&0.443&0.427&0.413&0.398&0.386&0.374&0.363\\
			$ \varPsi_B[\mathop{\mathbb{E}}(b^\ast)] $&218.78&226.07&235.00&248.91&266.01&292.70&330.51&377.20&440.95&528.11\\
			\hline	
		\end{tabular}
	\end{center}
\end{table}	
where both in Table \ref{LKBRAD} and \ref{LKRSRAD}, we find $ \mathop{\mathbb{E}}(C^\ast)=112.45 $ and $ \mathop{\mathbb{E}}(V_C)=150.00 $. So, 
\[
q=\frac{112.45}{150.00} = 0.7497.
\]

Thus, 
\[
F_{BC}=\frac{c-q\times 100}{q(200-100)} =\frac{c-74.97}{149.93-74.97}=\frac{c-74.97}{74.97},
\]
and the mean of this distribution is $ 112.45 $.
\begin{table}[H] 
	\small
	\begin{center}
		\caption{Brenda's optimal bids, probabilities of winning and her expected utilities when she is a risk-seeking bidder and she assumes that Charles is also a risk-seeking bidder} \label{LKBRSD}
		\begin{tabular}{ |p{2.65cm}||p{.88cm}|p{.88cm}|p{.88cm}|p{.88cm}|p{.88cm}|p{.88cm}|p{.88cm}|p{.88cm}|p{.88cm}|p{1cm}|}
			\hline 
			$r_B$ $ $&1.00&1.10&1.20&1.30&1.40&1.50&1.60&1.70&1.80&1.90\\
			\hline 
			\hline
			$\mathop{\mathbb{E}}(a_B)$ &1.00&1.09&1.17&1.26&1.34&1.43&1.52&1.60&1.68&1.76\\
			$ \mathop{\mathbb{E}}(b^\ast) $&100.79&98.73&96.90&95.19&93.63&92.13&90.76&89.53&88.38&87.38\\
			$ F_{BC}[\mathop{\mathbb{E}}(b^\ast)] $&0.954&0.914&0.879&0.846&0.815&0.786&0.760&0.736&0.714&0.694\\
			$ \varPsi_B[\mathop{\mathbb{E}}(b^\ast)] $&246.96&266.81&291.68&331.28&381.06&460.58&575.91&721.59&924.83&1208.68\\
			\hline	
		\end{tabular}
	\end{center}
\end{table}	

\begin{table}[H] 
	\small
	\begin{center}
		\caption{Brenda's optimal bids, probabilities of winning and her expected utilities when she is a risk-averse bidder and she assumes that Charles is a risk-seeking bidder} \label{LKRARSD}
		\begin{tabular}{ |p{2.65cm}||p{.88cm}|p{.88cm}|p{.88cm}|p{.88cm}|p{.88cm}|p{.88cm}|p{.88cm}|p{.88cm}|p{.88cm}|p{.88cm}|}
			\hline 
			$r_B$ $ $&0.10&0.20&0.30&0.40&0.50&0.60&0.70&0.80&0.90&1.00\\
			\hline 
			\hline
			$\mathop{\mathbb{E}}(a_B)$ &0.07&0.16&0.25&0.35&0.44&0.55&0.66&0.77&0.88&1.00\\
			$ \mathop{\mathbb{E}}(b^\ast) $&143.27&136.65&130.75&125.00&120.00&115.14&111.06&107.36&103.85&100.79\\
			$ F_{BC}[\mathop{\mathbb{E}}(b^\ast)] $&1.00&1.00&1.00&1.00&1.00&1.00&1.00&1.00&1.00&0.954\\
			$ \varPsi_B[\mathop{\mathbb{E}}(b^\ast)] $&201.14&201.51&202.09&203.08&204.47&207.05&211.21&217.99&229.14&246.96\\
			\hline	
		\end{tabular}
	\end{center}
\end{table}	
where both in Table \ref{LKBRSD} and \ref{LKRARSD}, we find $ \mathop{\mathbb{E}}(C^\ast)=77.36 $ and $ \mathop{\mathbb{E}}(V_C)=150.00 $. So, 
\[
q=\frac{77.36}{150.12} = 0.5157.
\]
Thus, 
\[
F_{BC}=\frac{c-q\times 100}{q(200-100)} =\frac{c-51.57}{103.15-51.57}=\frac{c-51.57}{51.57},
\]
$ $ and the mean of this distribution is $ 77.36 $.

\end{example}

\noindent Now, we provide a brief sketch of how Brenda would find the ARA solution when she wants to perform a level-$3$ analysis. In this case, \textit{k} $=3 $ and Brenda assumes that Charles is a level-$2$ thinker who would model Brenda as a level-$1$ thinker and believes that Brenda would model Charles as level-$0$ thinker.\\ 

To find the ARA solution in this case, Brenda would perform the level-$2$ analysis detailed above for Charles and obtain his optimal bid $C^\ast(W_{B})$ using \eqref{level KB}. Using \eqref{level KB} and using Monte Carlo methods to incorporate the uncertainty in $W_{B},$ she would get her belief about $ F_{BC} $, the distribution of Charles's (level-2 thinker) bid.  She can then obtain her optimal bid $b^\ast(w_{C})$ or the expected value of her optimal bid $\mathop{\mathbb{E}}(b^\ast)$ using similar process as in \eqref{nsp1} and \eqref{nsp2}, respectively.

\section{Summary and Further Work} \label{d}

In this paper, we propose a better way to model the FPSB auctions than what has previously been done. Specifically, we assume that the item being auctioned is a normal item. Many of the items typically auctioned using a FPSB auction are indeed normal items. First, we propose a new CRRA utility function that is realistic and constrains the bid value and true value in consideration with the wealth of the bidder, and a new CRRA parameter that models the change in risk behaviour of the bidder with increase in their wealth, as would be expected for a normal item. Secondly, we model the problem using the ARA approach and here we extend the ARA solution developed by \cite{banks2015adversarial} to consider the reserve price, to include not only risk- neutral but also, risk-averse and risk-seeking bidders and to allow each bidder to have different wealth.\\
Although ARA is also a Bayesian approach, one of the key advantages that ARA has over the Bayesian game theory is that because it does not aim to find an equilibrium solution for all the players, it is much easier to find an optimal solution even for complex problems. The second key advantage of ARA is that it allows the defender to model their adversary according to a variety of strategic solution concepts. In this paper we show how an ARA solution for an FPSB auction problem can be derived when assuming that the adversary is a non-strategic player and when assuming that they are a level-\textit{k} thinker instead. We provide numerical examples to illustrate these solutions. The solutions are very easy to be found using a basic Monte Carlo approach. The example shows that overall,  the optimal bids for risk averse bidders is higher than risk neutral bidders and that the optimal bids increase as the level of risk aversion increases. In contrast, but as expected, the optimal bids for the risk seeking bidders are lower than risk neutral bidders and the optimal bids decrease as the level of risk seeking behaviour intensifies. The examples also highlight that the probability of winning the bid increases with risk aversion and decreases with risk seeking behaviour. Further, it also shows that a bidder would typically bid higher with an increase in their relative wealth.\\
While it is possible for the defender to model the problem using ARA assuming various solution concepts for the adversary, it may be possible that the defender does not know how their adversary may solve the problem and therefore needs to incorporate concept (model) uncertainty into their solutions. This can be easily done. The defender first finds their optimal bid $\mathop{\mathbb{E}}(b^\ast)$ for each of the solution concepts that they wish to consider. The optimal bid derived for each solution concept $\mathop{\mathbb{E}}(b^\ast)$ is, in fact, the optimal bid conditional on the given model $\mathop{\mathbb{E}}(b^\ast | \mathcal{M}).$ The defender then has to elicit a probability distribution $p(\mathcal{M})$ that reflects their uncertainty on $\mathcal{M}.$ The optimal bid taking into account the model uncertainty is given by
\begin{equation}
 \mathop{\mathbb{E}}(b^\ast) = \sum_{\mathcal{M}} \mathop{\mathbb{E}}(b^\ast | \mathcal{M})p(\mathcal{M}).
\end{equation}

The practical challenge in adopting a Bayesian approach of incorporating uncertainties using prior distributions is the elicitation of these distributions. \cite{rios2016robustness} highlight the need and provide an outline for the robustness analysis for ARA. It is important to be able to to investigate the sensitivity of the optimal bid to any errors or mis-specifications in the utilities and the probabilities elicited for the analysis. Robustness analysis of ARA to these elicitations is necessary, but has yet to be developed.\\

In this paper, we derive the ARA solutions for the non-strategic play and the level-\textit{k} thinking solution concepts. However, the solutions for other solution concepts such as the BNE, mirror equilibrium or minimax approach have yet to be derived. Also, we have derived solutions for a two person game. General solutions for $n$-player games have to be derived too. Finally, other utility functions have also been proposed to model FPSB auctions, for example, the utility function that incorporates bidders winning and losing regret such as used by \cite{engelbrecht2007regret}. ARA solutions for a different utility function like this one have to be derived and sensitivity of the ARA solution to the choice of the utility function needs to be studied as well.

\bibliographystyle{plainnat}
\bibliography{references}
\end{document}